\documentclass{ws-p9-75x6-50}
\usepackage{amssymb}
\usepackage{epsfig}
\usepackage{psfrag}

\def\etal{{\it et~al.\/}}

\def\ccbeta{\eta}

\def\ltsim{\lesssim}

\def\CSS{{\rm CSS}}

\def\constant{{\rm constant}}

\begin{document}
\title{Episodic Self-Similarity in Critical Gravitational Collapse}
\author{J.~Thornburg, Ch.~Lechner, M.~P\"{u}rrer, and P.~C.~Aichelburg}
\address{Institut f\"{u}r Theoretische Physik, Universit\"{a}t Wien\\
	 Boltzmangasse 5, A-1090 Wien, Austria\\
   E-mail: jthorn@thp.univie.ac.at, lechner@thp.univie.ac.at,\\
	   mpuer@thp.univie.ac.at, pcaich@thp.univie.ac.at}

\author{S.~Husa}
\address{Department of Physics and Astronomy, University of Pittsburgh\\
	 3941 O'Hara Street, Pittsburgh PA 15260, USA\\
   E-mail: shusa@aei-potsdam.mpg.de}


%
%

\maketitle
\abstracts{
We report on a new behavior found in numerical simulations of spherically
symmetric gravitational collapse in self-gravitating \hbox{SU(2)}
$\sigma$~models at intermediate gravitational coupling constants:
The critical solution (between black hole formation and dispersion)
closely approximates the continuously self-similar (CSS) solution for
a finite time interval, then departs from this, and then returns to
CSS again.  This cycle repeats several times, each with a different
CSS accumulation point.  The critical solution is also approximately
discretely self-similar (DSS) throughout this whole process.}


\section{Introduction}

As summarized
in companion papers in these proceedings
   (Lechner \etal\cite{Lechner-etal:DICE-physics},
    Thornburg \etal\cite{Thornburg-etal:DICE-numerical}),
and described in detail elsewhere
   (Husa \etal\cite{Husa-etal:DICE-DSS-paper}),
we are studying critical phenomena in the \hbox{SU(2)} nonlinear
$\sigma$~model in spherical symmetry.  This model is parameterized
by a dimensionless coupling constant $\ccbeta$.  We denote the matter
field by $\phi = \phi(u,r)$, where $u$ is an outgoing null coordinate
(normalized to proper time at the origin) and $r$ is the areal radius.

This model is known to have a CSS solution for all $\ccbeta < 0.5$.
This solution can be explicitly constructed\cite{Bizon-Wasserman-2000-CSS},
and takes the form $\phi = \phi_\CSS(z; u_*)$, where $z = r/(u_*{-}u)$
and the (only) free parameter $u_*$ gives the location of the accumulation
point.

We consider a 1-parameter family of initial data, and fine-tune this
parameter so the initial data's evolution is very close to the
threshold of black hole formation.  At large (small) $\ccbeta$ the
evolution of such ``critical'' initial data is DSS (CSS) for a time,
before finally either dispersing or collapsing.

However, at intermediate $\ccbeta$ ($\approx 0.16$) a new behavior
appears, which we call ``episodic self-similarity'':
The field configuration closely approximates a CSS solution,
$\phi \approx \phi_\CSS(z;u_*^{(1)})$ in the inner part of the slice
for some finite time interval, then departs from CSS, and then returns
to closely approximate a CSS solution, $\phi \approx \phi_\CSS(z;u_*^{(2)})$
in the inner part of the slice for another finite time interval, then
departs, this cycle repeating several times.  The $u_*^{(k)}$ values
increase from one CSS episode to the next.

In addition, a large region of the evolution (spanning several CSS
episodes) is approximately DSS, but to a much lower degree of approximation
than the approximate CSS.

\begin{figure}[b]
\vskip-3.0mm
\begin{center}
\begin{psfrags}
\begin{picture}(80,80)
\psfrag{r}{\raise1.5mm\hbox{$r$}}
\psfrag{phi}{}
\psfrag{u=15.893  u*=16.193}
       {\hskip-2.0ex{\footnotesize $u{=}15.893$~~$u_*^{(1)}{=}16.193$}}
\psfrag{u=16.091}{\hskip-1.0ex{\footnotesize $u{=}16.091$}}
\psfrag{u=16.157}{\hskip-1.0ex{\footnotesize $u{=}16.157$}}
\psfrag{u=16.414  u*=16.662}
       {\hskip-2.0ex{\footnotesize $u{=}16.414$~~$u_*^{(2)}{=}16.662$}}
\put(-13,5){
	    \begin{picture}(0,0)
	    \put(-9,18){$\phi$}
	    \put(-9,58){$\phi$}
	    \put(-12.7,-8.9){\epsfig{file=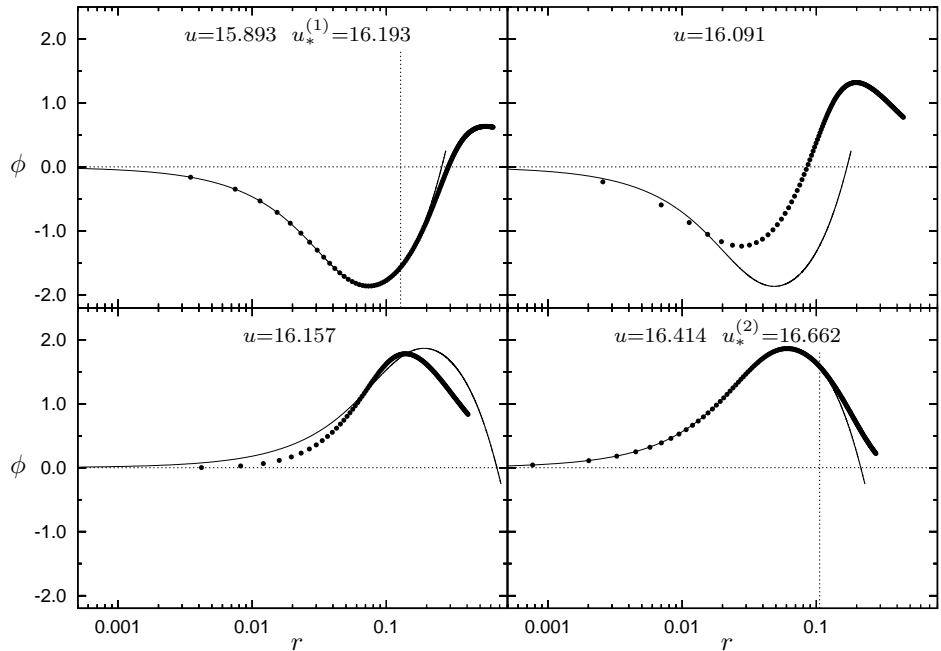}}
	    \end{picture}
	    }
\end{picture}
\end{psfrags}
\end{center}
\vskip-2.0mm
\caption[Critical Evolution and Fitted CSS Solution]
	{
	This figure shows selected $u = \constant$ slices in
	a numerical evolution of $\ccbeta = 0.16$ critical
	initial data ($\bullet$), with the best-fitting
	CSS solutions (computed independently for each slice)
	superimposed (---).
	For each slice where the fit is good and hence
	$u_* = u_*^{(k)}$ is well-defined, the vertical dashed
	line shows the CSS solution's self-similarity horizon.
	}
\label{fig-phi+phicss}
\end{figure}

Figure~\ref{fig-phi+phicss} shows an example of this behavior.
The field configuration never {\em exactly\/} matches a CSS solution,
but on the $u=15.893$ and $u=16.414$ slices (where the fit is good
and hence $u_* = u_*^{(k)}$ is well-defined),
$|\phi - \phi_\CSS| \ltsim 10^{-2}$ everywhere inside the self-similarity
horizon (the backwards light cone of the accumulation point $u_*^{(k)}$).
This region of the evolution is DSS to within $\sim 0.05$ in $\phi$.

We do not yet have a full understanding of episodic self-similarity
in terms of the standard phase-space model of self-similarity
\cite{Gundlach-critical-review}, but we think this behavior is caused
by a competition between nearby CSS and DSS attractors.


%
%


\end{document}